# Low frequency noise in nanoparticle-molecule networks and implications for in-materio reservoir computing.

Cécile Huez,[1] David Guérin,[1] Florence Volatron,[2]
Anna Proust[2] and Dominique Vuillaume.[1*]

1) Institute for Electronics Microelectronics and Nanotechnology (IEMN), CNRS, University of Lille, Av. Poincaré, Villeneuve d'Ascq, France.
2) Institut Parisien de Chimie Moléculaire (IPCM), CNRS, Sorbonne Université, 4 Place Jussieu, F-75005 Paris, France.

* Corresponding authors: dominique.vuillaume@iemn.fr

**Abstract.**

We study the low-frequency noise, *i.e.* flicker noise, also referred to as 1/f noise, in 2D networks of molecularly functionalized gold nanoparticles (NMN: nanoparticle-molecule network). We examine the noise behaviors of the NMN hosting alkyl chains (octanethiol), fatty acid oleic acids (oleylamine), redox molecule switches (polyoxometalate derivatives) or photo-isomerizable molecules (azobenzene derivatives) and we compare their 1/f noise behaviors. These noise metrics are used to evaluate which molecules are the best candidates to build in-materio reservoir computing molecular devices based on NMNs.

**Introduction.**

At the interface of nanostructures and bulk materials, macroscopic-scale nanostructures bridge the gap between the macroscopic and the nanoscopic material worlds. Among them, nanoparticles-molecules-networks (NMNs) are 2D arrays of molecularly functionalized nanoparticles connected on their periphery by several electrodes, which are used as a versatile platform to study the basic electron transport and optical properties in molecular electronics.[1-5] NMNs are also prone for several molecular electronics device applications, such as chemosensors, high sensitivity strain sensors, plasmonic devices, for instance.[4] Several studies have demonstrated their potentiality to implement devices for unconventional computing like reconfigurable logic gates,[6-10] neuro-inspired reservoir computing (RC).[10-12] Similar approaches were also developed with atomic contacts between the nanoparticles (or nanowires in several cases) instead of molecules.[13-18] The key features to implement physical reservoir computing devices are variability, strong nonlinear response and complex dynamic interactions inside the network.[19-21] The spatiotemporal dynamics inside the RC network generate noise that can be measured at the output electrodes. The low-frequency noise (LFN or flicker noise, also referred to as 1/f noise) has a power spectral density (PSD) that scales as $1/f^n$ with f the frequency and n the frequency exponent, which is usually between 1 and 2. The n=1 case is ubiquitous and it has been observed in a large number of systems (not only electronic devices). Albeit numerous and various physical mechanisms can be at its origin, it generally occurs in electronic devices as the consequence of fluctuations of charge carriers (number of carriers, mobility fluctuations) due to any source of carrier scattering.[22-25] A more specific case, n=2, is observed for two-level fluctuations, such as burst or random telegram signal (RTS) noise when the signal randomly fluctuates abruptly between two well-defined levels. In that case, the PSD has a Lorentzian shape with a $1/f^2$ dependence above a frequency corner (and a plateau below).[23, 24, 26] Thus, a value of n close to 2 is the fingerprint of additional and more complex noise sources in the system, likely favorable to an efficient physical RC. The two types of noise (as well as intermediate values of n) have been observed in molecular junctions (see reviews in Refs. 27, 28). In recent implementations of physical reservoir computing (RC) with various nanometarials and nanodevices, the relationship between LFN and the computational abilities of the RC was assessed with the objective to optimize the RC performances.[29, 30] While insufficient dynamics in nanomaterials or nanostructures used in physical RC can be compensated with additional external controls,[31] it is desirable to select a network with the highest complex dynamics.

Here, we compare the 1/f noise behaviors of the NMNs hosting alkyl chains (octanethiol), fatty acid oleic acids (oleylamine), redox molecule switches (polyoxometalate derivatives) or photo-isomerizable molecules (azobenzene derivatives). The choice of these four molecules is motivated by the following reasons. The first one (octanethiol) is a simple molecule used as a reference for our measurements by comparison with already reported results (alkyl chains).[32-35] The second molecule (oleylamine) is used to stabilize the starting Au NPs and it is interesting to know the LFN properties of the initial NMNs to assess the changes further introduced by more complex molecules of possible interest to implement a molecular RC. Polyoxometalates (POMs) were recently used, mixed with carbon nanotubes in a random network, to implement reservoir computing systems,[30, 36] while azobenzene derivatives are optically driven molecular switches[37-39] that were studied for reconfigurable logic circuits and reservoir computing approaches.[10] From the analysis of the 1/f noise, we conclude that highly dense NMNs with polyoxotungstates and NMNs with azobenzene-bithiophene in the *cis* isomer are the best candidates to build reservoir computing molecular devices based on NMNs.

**Results.**

Figure 1 shows an optical image of the 6-electrode connected NMN devices along with the 4 molecules used to functionalized the 7-8 nm in diameter gold nanoparticles (NPs). The NMN with oleylamine capped NPs is the precursor system for other functionalized NMNs derived by known ligand exchange protocols (see Methods and the Supporting Information). The 1-octanethiol chains are used as a reference sample since they are simple molecules and for comparison with already published results for NMNs functionalized with alkylthiols.[32-35] Then we studied the noise behaviors of NMN with $(TBA^+)_3[PW_{11}O_{40}(SiC_3H_6SH)_2]^{3-}$ ($PW_{11}SH$ or POM for short, $TBA^+$ is tetrabutylammonium, $[NBu_4]^+$) and azobenzene-bithiophene-butylthiol (azobenzene for short). The photo-switching behavior of the azobenzene molecules in a self-assembled monolayer molecular junction was previously reported with a conductance ratio up to *ca.* $7 \times 10^3$ between the "cis" isomer (high conductance state) and the "trans" isomer (low conductance state),[39] and a on/off conductance ratio up to *ca.* 600 in a NMN.[10, 40] Here we study the LFN behavior of the NMNs with the azobenzene in the two states. For the 4 molecules, we formed NMN with a roughly hexagonal arrangement of the functionalized AuNP between the 6 electrodes (see the experimental procedure for NMN fabrication in the Supporting Information and Figs. 1b-c and S2-S4 for their characterization by SEM). Image analysis (using ImageJ)[41] shows that the NP size and the inter-NP distances are Gaussian distributed with a mean NP diameter 7-8 nm (see Figs. S2-S4 in the Supporting Information). The inter-NP distance varies depending on the nature of the molecules, from ≈ 0.8 nm (for octanethiol-NMN) to ≈ 4.5 nm (azobenzene-NMN). Table 1 summarizes these structural characterizations.

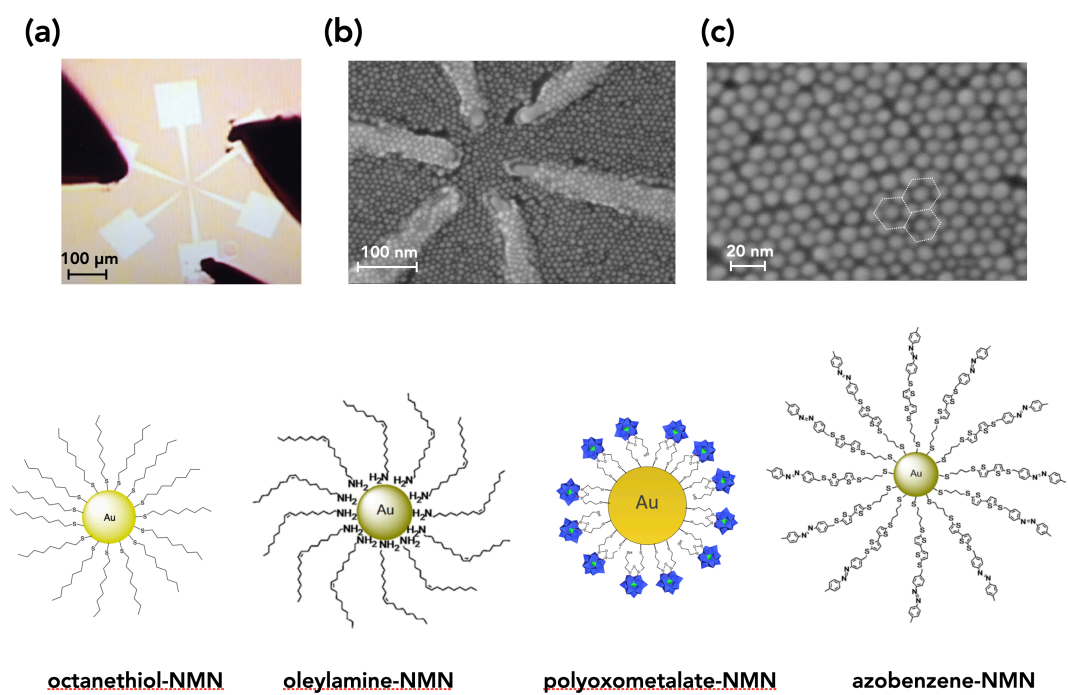

*Figure 1. (a) Optical image of the NMN devices with the 6 concentric electrodes and contacting pads (the black cones are the tips of the prober). (b and c) Scanning electron microscope images of the NMN at different magnification (214.64k and 659.75k, respectively). The panel (b) shows the 6 electrodes, the central ring between the electrode has a diameter of ca. 100 nm. Panel (c) is a zoom near the electrodes. The hexagonal packing of the functionalized NPs (here with POMs) is illustrated by the dotted white lines in the panel (c). Schemes of the four studied systems: NP functionalized with octanethiol, oleylamine, polyoxometalate and azobenzene, respectively (schemes not on scale). In the polyoxometalate case, the counterions (3 $TBA^+$ per POM) are omitted for clarity.*

|  | NPs diameter (nm) | NND (nm) | Molecule length (nm) |
|---|---|---|---|
| oleylamine-NMN | 7.8 ± 1.2 | 1.8 ± 0.4 | 2.0 [a] |
| octanethiol-NMN | 7.8 ± 2.7 | 1.5 ± 0.3 | 1.3 [b] |
| POM-NMN | 8.0 ± 1.2 (batch 1)<br>7.2 ± 1.6 (batch 2) | 1.4 ± 0.4 (batch 1)<br>2.1 ± 0.7 (batch 2) | 1.8 [c] |
| azobenzene-NMN | 9 ± 2 [d] | 4.5 [d] | 3.0 nm ("trans") [e]<br>2.5 nm ("cis") [e] |

*Table 1*. The SEM images (Figures S2-S4) were analyzed with ImageJ[41] to obtain the statistical distribution of the AuNP diameter. The inter-nanoparticle distance was calculated with a nearest neighbor distance (NND) ImageJ plugin. The histograms (Figures S2-S4) were fit with a Gaussian distribution, the mean values and standard deviations are given in this table. The molecule lengths are taken from the following references: [a] Ref. 42, [b] Ref. 43, [c] Refs. 44, 45, [d] Ref. 40, [e] Ref. 39.

The electrical measurements were carried out between two pairs of electrodes (PE) of the NMN randomly selected out of the 6 electrodes (Fig. 1b), see Methods. Figure 2a shows the typical current-voltage (I-V) curves measured between two PEs for a octanethiol-NMN and oleylamine-NMN (full data sets for different combinations of pairs of electrodes are shown in the Supporting Information, Fig. S5). The low-frequency noise was measured between the same two PEs and figure 2b shows the typical data for one PE of the octanethiol-NMN. The low-frequency noise is measured for applied voltages between 1.6 V and 11.2 V (by step of 1.6 V, applied using DC batteries, see Methods). The current power spectral density (PSD) $S_I(f)=\langle\delta I(t)^2\rangle/\Delta f$ follows a $1/f^n$ law (where $\langle\delta I(t)^2\rangle$ is the variance measured at a frequency f over a bandwidth $\Delta f$). The data sets for the other PEs of the octanethiol-NMN and the two PEs of the oleylamine-NMN are given in the Supporting Information (Fig. S6). The slope of the fits by a power law (straight lines in Fig. 2b) allows to determine the frequency exponent n versus the applied voltage (Fig. 2c). The values of n are not dependent on the applied voltage and close to n ≈ 0.9-1 for the octanethiol-NMN (average ⟨n⟩=0.93 ± 0.18) and ≈ 1.2 for the oleylamine-NMN (⟨n⟩=1.23 ± 0.10). The noise power is plotted versus the DC current (corresponding to the applied voltage according to the I-Vs) in Fig. 2d (The noise power is the PSD, $S_I(f)$, integrated over the scanned frequency range 1-100 Hz). We found the classical behavior that the noise power scales as $I^2$ (see discussion section).

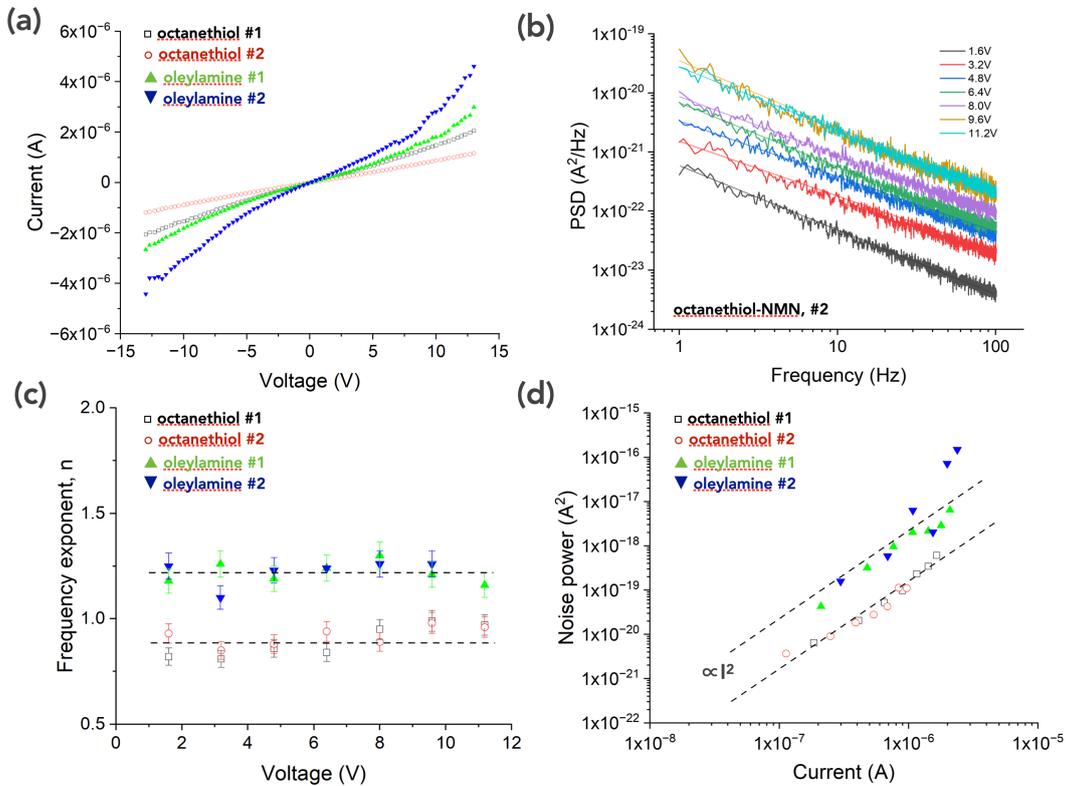

*Figure 2. (a) Current-voltage (I-V) curves for the two PEs (labels #1 and #2) for the octanethiol-NMN and oleylamine-NMN. (b) Current power spectral density (PSD) $S_I(f)$, versus frequency for the octanethiol-NMN, PE #2, measured at several applied voltages (straight lines are the fits by a power law). (c) Frequency exponent, n, versus the applied voltage. (d) Noise power (i.e. the PSD, $S_I(f)$, integrated over the scanned frequency range 1-100 Hz) versus the DC current,*

*which is taken from the I-V shown in the panel (a). The dashed lines are a guide for the eyes showing the $I^2$ scaling.*

Figures 3a-b show the I-V data set recorded for 3 different POM-NMNs (from 2 batches, see the Supporting Information) between several PEs. The three NMNs clearly differ by their level of currents, which we refer to as high current (HC, Fig. 3a), medium current (MC) and low current (LC), Fig. 3b and Fig. S7. The POM-NMN with high current belongs to batch 1, while the MC and LC devices correspond to two NMNs in the same chip from batch 2. The I-Vs of the HC NMNs are systematically characterized by large current instabilities for voltages (in absolute values) between *ca.* 2 and 10 V. The I-Vs for the MC and LC NMNs are more stable, we note small fluctuations at $|V| \gtrsim 10V$ for the MC NMN, while the I-Vs of the LC NMN are fully stable (Fig.S7). The low-frequency noise was measured for 2 PEs of each NMNs (in the following referred to as POM #1 and #2 for the HC NMN of batch 1, POM #3 and #4 for the MC NMN of batch 2 and POM #5 and #6 for the LC NMN). Figure 3c gives the frequency exponent, n, for these samples (with the corresponding PSD data sets in the Supporting Information, Fig. S8). The main feature is higher n values ($\approx$ 1.2-1.7, $\langle n \rangle$=1.46 ± 0.21) for the HC NMN compared to the MC and LC NMNs for which the n values are indiscernible (randomly dispersed in the range n ≈ 0.9-1.2, $\langle n \rangle$=1.09 ± 0.10). A difference is also notable in the noise power versus current behaviors (Fig. 3d). While the MC and LC NMNs follow the usual $I^2$ dependence, the noise powers for the HC NMNs are randomly dispersed with higher values.

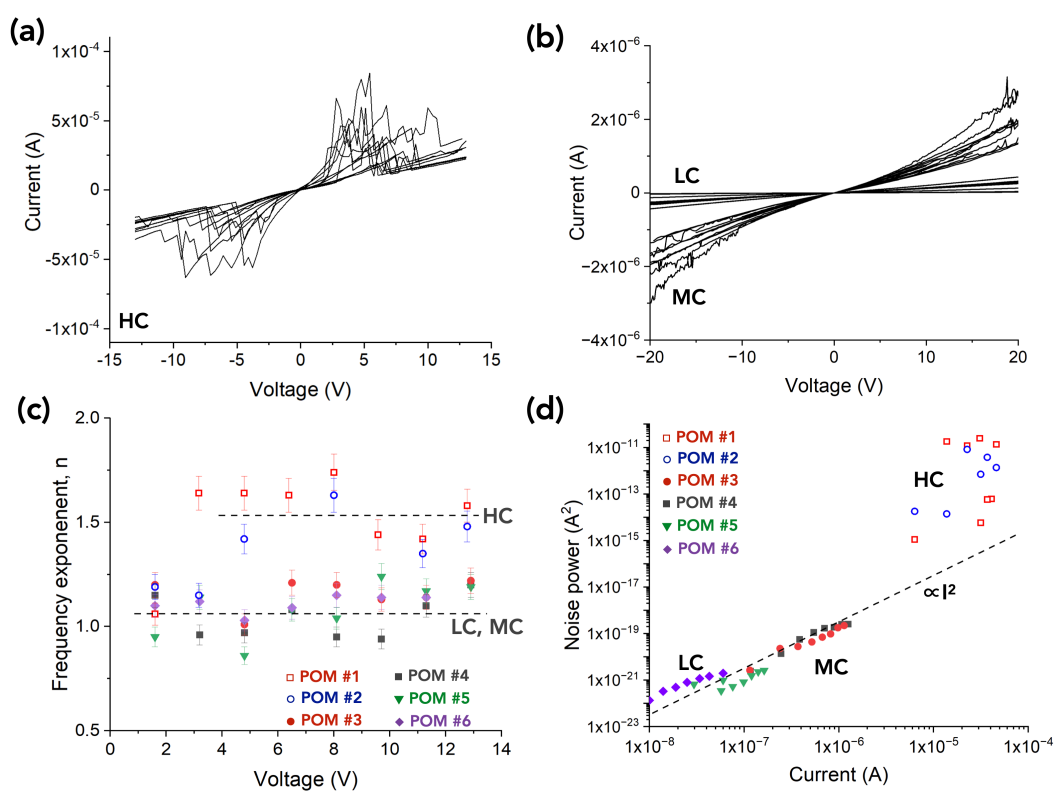

*Figure 3*. *(a)* Current-voltage (I-V) curves for the 12 independent PEs for the POM-NMN of batch 1. *(b)* Current-voltage (I-V) curves for the 24 independent PEs for two POM-NMNs of batch 2 (14 for the MC group and 10 for the LC group). *(c)* Frequency exponent, n, versus the applied voltage for 2 PEs, POM #1 and #2, for the NMN of batch 1, POM #3 and #4 for one NMN of batch 2 and POM #5 and #6 for the second one with the lowest currents. *(d)* Noise power versus the DC current, which is taken from the I-V shown in the panel (a). The dashed line is a guide for the eyes showing the $I^2$ scaling. Same symbols as in panel (c).

The same set of measurements for the azobenzene-NMN devices is shown in Fig. 4 for the azobenzene-NMNs with the molecules in their *trans*-state (initial measurements) and their *cis* state after UV light illumination (at 365 nm for 1 h). In that case, we have measured rigorously the same PE of the azobenzene-NMN before and after photoisomerization to assess the change due to the azobenzene isomerization in the NMN. After the UV illumination, we clearly observed an increase of the current by a factor ≈ 15 (Fig. 4a) in agreement with our previous finding that the *cis*-azobenzene NMNs are more conducting than the *trans*-azobenzene NMNs.[10, 40] From the $1/f^n$ measurements (Figs. 4b), we show that the frequency exponent, n, has different behaviors for the two azobenzene isomers. For the *trans*-state, n increases from ca. 1 to 1.3 when increasing the applied voltage from 1 to 12 V, while for the *cis* state, it is constant to n ≈ 1.4 (⟨n⟩=1.37 ± 0.05, Fig. 4c) at all the applied voltages. The noise power scales as $I^2$ (Fig. 4d).

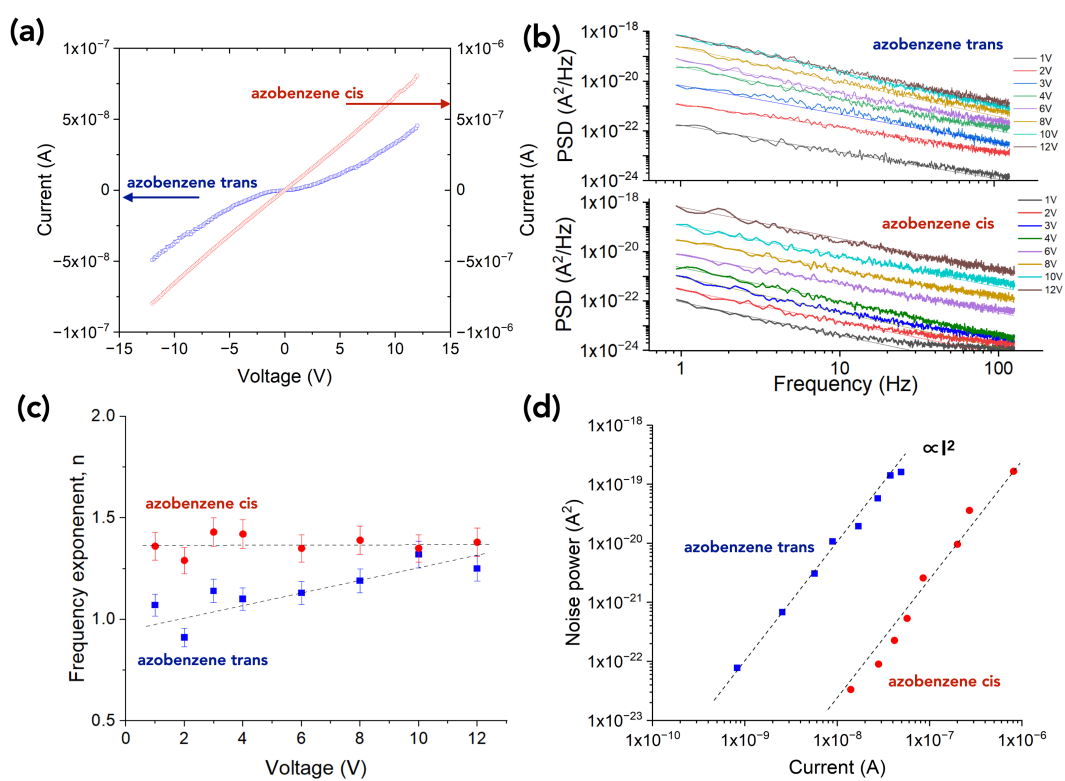

***Figure 4**. (a) Current-voltage (I-V) curves for the same PE with the azobenzene molecules in the trans- and cis-states. **(b)** Current power spectral density (PSD) $S_I(f)$, versus frequency measured at several applied voltages and for the two states of the azobenzene molecules. **(c)** Frequency exponent, n, versus the applied*

*voltage. **(d)** Noise power versus the DC current, which is taken from the I-V shown in the panel (a). The dashed lines are a guide for the eyes showing the $I^2$ scaling.*

## Discussion.

Since the source of noise in electronic devices are multiple,[25, 46] we discuss several possible physical origins focusing on the more relevant for molecular based devices.[27, 28]

***Octanethiol and oleylamine NMNs.*** The LFN behavior of the NMNs follows the empirical Hooge law given by[22, 25]

$$S_I(f) = \frac{\alpha_H I^2}{N f^n} \quad (1)$$

where I is the DC current, f the frequency (with a frequency exponent, n), N is the number of free carriers in the device and $\alpha_H$ the Hooge constant. The LFN behavior of the octanethiol-NMN is in agreement with previous reports for similar systems (NPs capped with alkylthiols of 8 to 12 carbon atoms)[32, 33] with n close to ≈ 1 (⟨n⟩=0.93 ± 0.18) (Fig. 2), independent of the applied voltage (or current passing through the NMNs). The frequency exponent is slightly larger, n ≈ 1.2, for the oleylamine-NMNs (⟨n⟩=1.23 ± 0.10). We also note a larger noise power in the latter case. The LFN can originate from various physical phenomena, such as conformational fluctuations of the molecules, fluctuations of the molecule-metal bonding, tiny motion of the NPs. It has been shown that alkyl chains have different electrical conductances whether they are in their "all trans" conformation or whether they have "gauche" defects along the chain.[47] For a long chain as oleylamine, the probability to have "gauche" defects is higher than for a shorter one as octanethiol, and thus should induce more noise fluctuations. However, in the two cases, the nearest neighbor distance is almost equal to the molecule size (NDD in Table 1) and consequently they are likely strongly interdigitated inside the NP gaps, a feature that probably minimizes this noise source. The other difference is the chemical nature of the anchoring groups (thiol versus amine). From the study of single molecule junctions (STM break junction) and DFT calculations, we know that, at low voltages (typically few tens of mV), the amine anchoring group leads to a less dispersed conductance data set than the thiol-Au junctions,[48] thus potentially smaller fluctuations and current noise. This feature is due to the flexibility of the coupling of the N lone pair to Au. However, at higher voltages (as it is the case in NMNs), it has also been observed that the I-Vs of molecular junctions using amine anchoring groups exhibit more instability and noise than the ones with S-Au anchors. This is due to the weaker binding strengths of amine to gold.[49] However, these subtle voltage-dependent noise behaviors, observed in single-molecule experiments,[48, 49] were not observed in our SAM-based devices, where larger n and noise power were always measured for oleylamine-NMNs irrespective of the applied voltage (Figs. 2c and 2d). We suggest that the larger noise in the oleylamine-NMNs is mainly due to the molecule-metal interface bonding rather than by molecular conformational fluctuations of the molecule backbone itself. We notice that the oleylamine-NMNs have a slightly higher level of current than the octanethiol-NMNs (Fig 2a), albeit the nearest neighbor distance is larger (NDD, Table 1), which should have induced a less efficient electron transport from one NP to the next one through the molecules, according to the well-established exponential decrease of the off-resonant electron transport in aliphatic chain molecular junctions.[50, 51] This apparent contradiction is understood because the NPs network in the octanethiol-NMNs is more disordered, with a lower average density of NPs and more voids (see Figs. S2-S3). Consequently, the number of electron conduction pathways between the PEs though the network is reduced (compared to the better organized oleylamine-NMNs), which can counterbalance the higher electron transport of the individual NP-molecules-NP due to the smaller nearest neighbor distance for the octanethiol-NMNs.

***Polyoxometalate NMNs.*** The I-V stability and LFN behaviors of the POM-NMNs depend on the level of DC current passing through the NMNs (Fig. 3). The larger currents for HC NMNs (from batch 1) is consistent with the smaller nearest neighbor distance (Table 1) and thus a better electron transfer between adjacent NPs through the molecules. The MC and LC NMNs (from batch 2) have a larger nearest neighbor distance (Table 1). The difference between the MC and LC NMNs (two NMNs randomly selected on the batch 2 chip) can be explained considering the large dispersion of the NND (Table 1). For the HC NMN, large instabilities of the I-V traces are systematically observed in all the I-V traces for applied voltages ≳ 2 V (in absolute value) and ≲ 10 V. These sudden and random instabilities during the voltage sweep translate into burst or RTS (random telegraph signal) noise in the time domain and they are likely the origin of the larger frequency exponent (⟨n⟩=1.46 ± 0.21, Fig. 3). A pure RTS noise corresponds to a Lorentzian PSD given by[26]

$$S_I(f) \approx \frac{4I^2\tau}{1+(2\pi f\tau)^2} \qquad (2)$$

(i.e. a PSD ∝ $1/f^2$ above a corner frequency that depends on the time constant τ of the fluctuations). If the amplitude of the Lorentzian noise is of the same order as the strict flicker noise, a mix of the two can result to the observed $1/f^{1.4-1.5}$ behavior, at least for a certain frequency window. The $I^2$ behavior of noise power is no longer observed for the HC POM-NMNs (Fig. 3d). A critical current of few µA seems the condition to observe this RTS-like noise behavior with high n and large noise power (Fig. 3d). Indeed, this behavior clearly disappears when lower currents (< 0.3 µA) are passing through these HC NMNs (at low applied voltages ≤ 1 V) - Fig. S9. In these conditions, stable I-Vs are recovered, and the LFN is characterized by a frequency exponent n ≈ 1.1-1.2 (Fig. S9). Similarly, the LFN of the MC and LC NMNs (for which DC currents are ≤ 2 µA, Figs. 3b and 3d) also display the usual "1/f" behavior with ⟨n⟩=1.1 ± 0.1 (Fig. 3c) and the noise power scaling as $I^2$ (Fig. 3d). Since the same voltages are applied on all the NMNs, this change of the LFN behavior is current-driven, with a critical current ≳ 5 µA to induce this $1/f^{1.4-1.5}$ noise behavior (Fig. 3d). We suggest that this RTS-like fluctuations are due to trapping/detrapping of electrons by the POMs, which are known as efficient electron attractors.[52-54] When the electron flux though the NMN is high enough (as in HC NMNs), POMs can capture a significant number of electrons and get reduced. It is known that the electrical conductance of reduced POMs (in thin film, in self-assembled monolayer junction, as well as in single molecule junction)[52, 55, 56] is increased (mainly due to a lowering of the lowest unoccupied molecular orbital, LUMO). Thus, the conductance of the NP-POM-NP building blocks in the NMNs increases, and so do the global current through the NMNs. Simultaneously, the POMs become negatively charged (the number of countercations is not changed) and a strong Coulomb repulsion between the closely adjacent POMs can force detrapping of electrons from the POMs to equilibrate the electrostatic landscape in the NMNs. This trapping/detrapping dynamics leads to the RTS-like noise. When the current in the NMNs is too low, less electrons are trapped in the POMs and the competition between trapping and Coulomb detrapping is reduced. The RTS-like noise is consequently reduced in the NMNs and only flicker noise is observed in MC and LC NMNs. The voltage window (ca. 2<|V|<10 V) where the large fluctuations are observed in the I-V curves of the HC NMNs (Fig. 3a) correspond roughly to a local voltage of 0.2 - 1 V inside a single NP-POMs-NP (considering an average number of 10 single junctions between the peripheral electrodes of the NMNs, see details in the Supporting Information). The LUMO of the POM is readily accessible within this voltage range. The redox potential (-0.43 V/SCE in acetonitrile)[57] gives a LUMO at ≈ -4.25 eV from vacuum, and considering a work function of ≈ -4.8 to -5.1 eV for the gold NPs, the LUMO is at ≈ 0.55 - 0.85 eV from the Au Fermi energy (or even smaller, since the WF of gold NPs must be weaker, especially depending on the charge states of the NPs).[58]

Since the $1/f^{1.4-1.5}$ behavior is only observed in HC NMNs, another noise contribution could come from the current crowding[59] at the nanoscale molecule-NP contacts and/or at the contact between the functionalized NPs and the nano-electrodes. Current crowding effect unavoidably appears at high local current density due to resistance mismatches and scattering of charge injections through the nanoscale contact. This current crowding effect can also increase the noise as observed in various devices.[46, 60, 61]

Finally, we also note that n ≈ 1.4-2 was observed for single molecule junctions (STM-break junction and mechanically controlled break junction) and was tentatively ascribed to conductance fluctuations due to atomic rearrangements at the molecule/metal interface.[62, 63]

**Azobenzene NMNs.** For the azobenzene-NMNs (*cis* state), we observed a frequency exponent at around 1.4 (⟨n⟩=1.37 ± 0.05, Fig. 4). For these NMNs, the DC current is weak and the current crowding effect can be ruled out. Based on our previous molecular simulations,[40] we have demonstrated that the *cis*-azobenzene molecules in the NP-NP gap have larger conformational fluctuations than the *trans*-azobenzene, because the *cis*-azobenzenes are weakly interdigitated in the gap (or even having completely lost contact with each other), with low interactions between them. On the contrary, with the azobenzene in the *trans* isomer and more molecules interdigitated, the intermolecular interactions are increased, given rise to a more stable NP-molecules-NP building block structure in the NMNs. This picture is consistent with the smaller n value for the *trans* azobenzene-NMNs (Fig. 4). In this latter case, the striking feature is the increase of the frequency exponent n with the applied voltage from ca. 1 to 1.3. It is known that the *trans-cis* isomerization of azobenzene molecules can also be electrically induced by applied an electric field of 0.1-0.7 V/Å.[64, 65] We can hypothesize that such an electrically induced *trans-cis* isomerization can stochastically happen in the NMNs, inducing more conductance fluctuations and given rise to the observed increase of the frequency exponent n as increasing the applied voltages. However, in the voltage range 8-12 V (voltages at which the LFN behavior of the *trans* azobenzene-NMNs tends to be similar to the one of the *cis* azobenzene-NMNs), the local electric field across an individual NP–molecules–NP building-block junction in the NMNs is ≈ 1.2 - 5.3x10$^{-2}$ V/Å (see details in the Supporting Information), far below the electric field applied in the STM experiments.[64, 65] We conclude that this electrically induced isomerization can be ruled out in our case.

Another possible physical origin of this voltage-dependent increase of noise is the interactions with vibrational modes of the azobenzene molecules. Inelastic electron tunneling spectroscopy experiments supported by theoretical (DFT) studies on single azobenzene molecules (by mechanical-controlled break junction) have revealed the existence of many vibrational modes in the energy range 0-0.2 eV, especially with a strong mode at ≈ 0.18 eV for the *trans* isomer and ≈ 0.2 eV for the *cis* one.[66] These vibrational modes induce inelastic scattering in the electron transport through the molecular junctions and thus conductance fluctuations and noise. Increasing the voltage across the individual NP-molecule-NP in the NMN (typically up to ≈ 1 V at an external applied bias of 10 V, see the Supporting Information), more and more vibrational modes can interact with the traveling electrons in the molecular junction, increasing the number of conductance fluctuations and the noise feature in the overall NMN. Here, this effect is only observed for the *trans*-azobenzene-NMNs since the less stable *cis*-azobenzene-NMNs are already the subject of a larger noise (*vide supra*), which can hide this voltage-dependent noise effect.

### *Implications for physical RC*.

A reservoir computing (RC) is a type of simplified recurrent neural network in which the training to compute a given task is reduced compared to other recurrent neural networks. The key part of RC is the reservoir, a randomly connected network of nodes and links, featuring large variability, strong non-linear responses and complex dynamics (see more details in the Supporting Information).[19, 20, 67-69] High computing capacities require complex non-linear interactions, which at the same time are also sources of noise in these systems. The existence of correlations between 1/f noise in a complex system and the ability of this latter to perform efficient information processing is an open question, both in biological systems and in artificial man-made neuromorphic computing.[29, 70-74] Albeit it was suggested how to overcome insufficient dynamics in emergent nanomaterials or nanostructures used for the implementation of physical RC using additional external controls,[31] it is desirable to select systems with the highest complex dynamics. On the other hand, very noisy, chaotic systems have also been suggested for computing.[75-78]

Some works have recently examined the relationship between LFN and the computational abilities of the RC, with the objective to optimize the RC performances.[29, 30] In the following, they are briefly discussed as a basis to assess which of the four molecule-NMNs studied here are the most appealing for a possible RC physical implementation.

In dopant atom networks in silicon, Chen et al.[29] have observed that the neuromorphic computational ability is optimized when the device is biased in a narrow voltage range (around 0.4 V in this case) for which the noise exponent, n, rises from 0.2 to 0.8 for this specific device. This condition also coincides with a peak in the signal-to-noise ratio (SNR). This noise behavior is related to the charge carrier transport mechanism (hopping conduction in the impurity band of the doped silicon) and dynamic rearrangements of clusters of charge carriers around the dopant atoms. In the same voltage range, the SNR goes to a maximum because the signal and noise scale differently with the applied voltage (or the current passing in the device). The signal scales sub-linearly with the current, while the noise scales quadratically with the current.[29] As such, the reported behavior of these devices is likely specific to this system and not easily extrapolated to others emergent technologies (discussed below). In particular, we note that in these devices the maximum of n is about 1 because the devices were made with a modern silicon technology, for which noises sources are well controlled. We also note that the experiments in Ref 29 were done at 77K (to access the variable range hopping transport regime) and it is difficult to compare with the other systems discussed below and with our work (all done at room temperature).

Dense and random networks of carbon nanotubes (CNT) complexed with POMs were used to implement RC.[36] In a benchmark task of object classification, it was observed that the success of the classification tasks is correlated with the presence of LFN at the outputs of the CNT/POM reservoir (*i.e.* the task is not processed in the presence of white noise, or LFN with n < 0.2).[79] A frequency exponent, n, between 1.2 and 1.5 is required in that case. This LFN was associated to stochastic changes in the current-voltage characteristics of the CNT/POM basic building blocks of the RC, likely coming from redox switching of the POMs upon charge injection and accumulation in the network.[36]

Atomic switch networks, made of random arrays of sulfurized silver ($Ag_2S$) nanowires have also demonstrated RC capabilities.[15, 80] These computing abilities rely on a dense network of interconnected atomic metallic filaments that develop at the crossing of nanowires. These devices provide multi-state conductance values, strong non-linearity and large variability of the DC and time dynamic properties, all features necessary for an efficient RC system.[19, 20, 67-69] These non-linear dynamics generate fluctuations and noises. It was observed that atomic switch networks without LFN noise (e.g. made of un-sulfurize Ag nanowires, showing only white noise) are not capable of performing neuromorphic tasks as do the networks of silver sulfide nanowires, which display $1/f^n$ noise, with n ≈ 1.4.[80]

For the RC implemented with emergent technologies like the CNT/POM devices and the atomic switch networks, the best computing performances correspond to the largest noise exponent n observed for these two systems (*vide supra*). Albeit being far to be generalizable and considered as a universal relationship, we suggest that the same conclusions hold in our approach with nanoparticle/POM. In addition, in all the three cases (CNT/POM, atomic switch networks and POM-NMN) the high n (1.4-1.5) is correlated with stochastic events that induce abrupt changes in the current-voltage curves (*e.g.* see Figs. 1a and 3c in Ref 36, Fig. 3 in Ref 80) as also observed for polyoxometalate-NMNs in our work (Fig. 3a), albeit their physical origins are not strictly similar (redox switching for POMs in Ref. 36 and in our work, formation/breaking of atomic filaments/contacts in Refs 15, 80).

     All the NMNs studied here show a large degree of variability as reflected by the dispersion of the I-V traces recorded between randomly selected PEs (*e.g.* Figs. S5 and S7). Considering a high value of the frequency exponent, n, as a fingerprint of a complex dynamic behavior[26] whatever the physical origin of this noise, we suggest that the best candidates are POM-NMNs with a relatively high level of current (> few µA) and the azobenzene-NMNs with the azobenzene molecules in the *cis* isomer. These two systems also satisfy the condition of the nonlinearity of the NMN responses. We have previously demonstrated that *cis* azobenzene-NMNs are characterized by a rich high harmonic generation, including harmonic distortion, interharmonic distortion and intermodulation distortion, as a result of complex nonlinear interactions of electron transport in such a highly connected and recurrent networks of molecular junctions.[10] A similar high harmonic generation behavior is also observed for the POM-NMNs (Fig. S10 in the Supporting Information). Moreover, it is also benefit to be able to tune the frequency exponent, n, (through the applied voltage or injected current, e.g. from 0.8 to 2.1 in Ref. 81) to further move towards complex adaptive systems[81] and task-adaptive RC.[82] Such a behavior was clearly observed for the *trans*-azobenzene-NMNs (Fig. 4c) and POM-NMNs in the low bias regime (Fig. S9c). Thus, we conclude that POM-NMNs and azobenzene-NMNs are suitable as

physical reservoir computing. Finally, we note that fading memory (another characteristic useful in neuromorphic and RC systems to generate short-term and/or long-term plasticity) is also present in the azobenzene-NMNs and POM-NMNs. Indeed, the reduced POM state and the *cis*-azobenzene isomers are known to be metastable states, spontaneously returning (slowly) to their more energy stable states (neutral POM, *trans*-isomer)[10, 39, 40, 56] and, thus, they are likely participating as sources of fluctuations and complex dynamics in these NMNs. It is clear that all these studies call for more investigations to turn hypothesis in sound correlation between noise and reservoir computing efficiency.

**Conclusions**

The low-frequency noise, $1/f^n$ noise, of nanoparticle-molecule-networks (NMNs) with octanethiols and oleylamines obeys the Hooge law with a frequency exponent n close to 1 and 1.2, respectively, independent of the applied voltage. The noise amplitude scales quadratically with the DC current passing through the NMNs. The slight difference between the two NMNs is ascribed to the different molecule/gold bonding (thiol vs. amine groups). A larger frequency exponent, n ≈ 1.4-1.5 is observed for the NMNs with polyoxotungstates and *cis* isomer azobenzenes. These higher values are due to the presence of larger conductance fluctuations, like random telegraph signal, in these two NMNs. In the case of the azobenzene NMNs, this additional noise is likely due to larger molecular structure fluctuations than for the *trans* isomer. In this latter case, the frequency exponent increases (from ≈ 1 to 1.3) with the voltage. We interpret this behavior as a consequence of the inelastic scattering of electron transport by the azobenzene vibrational modes. In the case of polyoxotungstates NMNs, we propose a current-driven stochastic redox switching of the molecules inserted between the gold nanoparticles. For the less dense NMNs with lower current, the low-frequency noise recovers the usual flicker noise with n close to 1. From these results, we conclude that the polyoxotungstate NMNs with the highest density of nanoparticles and NMNs with the azobenzene in the *cis* isomer are the most suitable for the implementation of in-materio reservoir computing devices.

# Methods

*Molecule synthesis.*

1-octanethiol and (9Z)-Octadec-9-en-1-amine (oleylamine for short) were purchased from Aldrich and used as received. The azobenzene derivatives were synthesized as reported in our previous work.[10, 39, 40, 83] The synthesis and characterization of $PW_{11}SH$ derivatives have been reported elsewhere.[44, 57]

*Nanoparticle Molecule Network fabrication .* We started with the synthesis of oleylamine-capped Au NPs as previously reported (more details in the Supporting Information).[84] The NMNs were deposited on $Si/SiO_2$(200 nm thick) substrate equipped with the electrodes (fabricated by e-beam and liftoff, 2 nm thick Ti anchoring layer and 12 nm thick Au) by a Langmuir and transfer method according to Santhanam et al.[85] Then, we performed a ligand exchange to functionalize the NMNs with the molecules of interest (octanethiol, azobenzene and polyoxometalate). The fabrication of A azobenzene-NMNs was already reported and fully characterized in our previous work.[10, 40] The full fabrication details for the octanethiol-NMNs and polyoxometalate-NMNs are given in the Supporting Information.

*Physical characterizations of the NMNs.*

Scanning Electron Microscopy (Zeiss ULTRA55) was used to inspect the electrodes and NMNs (electron beam 10kV). The SEM images of NMNs were treated with ImageJ[41] to measure the size of the NPs and the inter-nanoparticle distance using the function analyze particles and the plugin NND (nearest neighbor diameter).

*Electrical measurements.*

The NMNs were electrically connected with a micromanipulator probe station (Suss Microtec PM-5) installed inside a glovebox (MBraun, nitrogen filled <0.1 ppm of oxygen and water vapor) to avoid any degradation of the molecules. The current-voltage (I-V) curves were acquired with an Agilent B2901A SMU (source/measurement unit). The I-V curves were acquired in a quasi-static mode with a DC voltage following a staircase ramp with a voltage step of 0.1 V, a step time of 1 s resulting in a low sweep rate of 0.1 V/s. All the I-V curves were recorded in the upward direction from negative (-13 or -20 V) to positive voltages (13 or 20 V). For the noise measurements, a DC bias is applied on one of the NMN electrodes by a series of 1.6 V battery or an ultra low-noise DC source (Shibasoku PA15A1 or Yokagawa 7651), the output currents are simultaneously measured on two other electrodes of the NMN by two trans-impedance amplifiers (model1211 from DL Instruments or Stanford Research Systems SR570) and analyzed by a two-channel digital signal analyzer (Agilent 35670A) in the frequency or time domains (for the $1/f^n$ noise, the reported PSD is an average done on 50 scans from 1 to 100 Hz). The noise floor of the setup is $\approx 10^{-28}$ $A^2$/Hz (measured with no sample, the prober tips raised, Fig. S11). The thermal noise of the samples is in the range $\approx 10^{-29}$ to $10^{-27}$ $A^2$/Hz depending on the sample resistance R according to $S_{I,th}=4kT/R$ (k the Boltzmann constant, T the temperature). To induce the *trans-to-cis* photoisomerization of the azobenzene we used a UV lamp (UVP-3UV from Analytik Jena) at a wavelength at 365 nm and a power of *ca.* 0.5mW/cm² placed at a distance of *ca.* 1cm from the sample.

## Online content

Details on the NMN fabrication, scanning electron microscope images and nanoparticle size analysis, additional current-voltage and power spectral density data, high harmonic generation are available at....

## Author Contributions

C.H. and D.G. fabricated the NMNs. F.V. and A.P. synthesized and characterized the POM. C.H. and D.V. carried out the electrical measurements. D.V. and A.P. conceived the project, supervised it and secured the project funding. The manuscript was written by D.V. with the contributions and comments of all the authors. All authors have given approval of the final version of the manuscript.

## Conflicts of interest

There are no conflicts to declare.

## Data availability

The data supporting this article are available from the corresponding author upon reasonable request.

## Acknowledgements.

We acknowledge support of the CNRS, project "neuroPOM", under a grant of the 80PRIME program. We thank C. Boyaval (IEMN) for the SEM images.

## References.

# SUPPORTING INFORMATION

# Low frequency noise in nanoparticle-molecule networks and implications for in-materio reservoir computing.

Cécile Huez,[1] David Guérin,[1] Florence Volatron,[2]
Anna Proust[2] and Dominique Vuillaume.[1*]

1) *Institute for Electronics Microelectronics and Nanotechnology (IEMN), CNRS, University of Lille, Av. Poincaré, Villeneuve d'Ascq, France.*
2) *Institut Parisien de Chimie Moléculaire (IPCM), CNRS, Sorbonne Université, 4 Place Jussieu, F-75005 Paris, France.*

\* Corresponding authors: dominique.vuillaume@iemn.fr

## NMN fabrication.

***Electrodes on Si/SiO$_2$.*** We used a ⟨100⟩ oriented silicon wafer covered with a 200 nm thick silicon dioxide thermally grown at 1100 °C during 135 min in a dry oxygen flow (2 L/min) and followed by a postoxidation annealing at 900 °C during 30 min under a nitrogen flow (2 L/min) to reduce the density of defects into the oxide and at the Si/SiO$_2$ interface. The metal electrodes were fabricated by e-beam lithography. We used a 45 nm-thick PMMA (4% 950 K, diluted with anisole with a 5:3 ratio), with an acceleration voltage of 100 keV and an optimized electron beam dose of 370 µC/cm$^2$ for the writing. After the resist development (MIBK:IPA 1:3 during 1 min and rinsed with IPA), a metallic layer (2 nm of titanium and 12 nm of gold) was deposited by e-beam evaporation followed by the liftoff process using remover SVCTM14 during 5 h at 80 °C. We obtained well defined 6 coplanar electrodes arranged around a ring with a diameter between 80 to 120 nm.

**_Synthesis of molecularly functionalized Au NPs and deposition on a substrate._**
To better control the size of gold nanoparticles (≈10 nm), we decided to prepare oleylamine-coated Au NPs by a phase transfer protocol[1,2] from citrate-coated AuNPs instead of the direct reduction[3,4] of tetrachloroauric salt with oleylamine (Fig. S1). First of all, a 100 mL aqueous solution of 10 nm citrate-AuNPs was obtained following the Turkevich method.[5] A solution with 1 mL of tetrachloroauric acid trihydrate $HAuCl_4·3H_2O$ (1%) in 79 mL of deionized water was prepared. Then a 20 mL reducing solution with 4 mL of trisodium citrate dihydrate (1%) and 80 μL of tannic acid (1%) in 16 mL of deionized water was added rapidly to the Au solution under vigorous stirring (important : both solutions were mixed at 60 °C). The mixture was boiled for 10 min before being cooled down to room temperature. A continuous stirring was applied throughout the process. Then, the 100 mL solution of citrate capped NPs was extracted with 20 mL of hexane containing 0.2 mL of oleylamine. After vigorous stirring, in a separatory funnel, the organic phase was isolated and washed twice with deionized water. The dark red suspension was distributed in centrifuge tubes and then added with 50 to 70% ethanol until the beginning of the agglomeration (purple shift). After centrifugation at 7000 rpm for 5 min, the precipitate was washed with absolute ethanol then redispersed in hexane. The washing of the NPs by precipitation with ethanol then redispersion in hexane was repeated twice in order to eliminate the excess of oleylamine. The NPs suspension is stable in hexane or toluene. It is stored in the refrigerator.

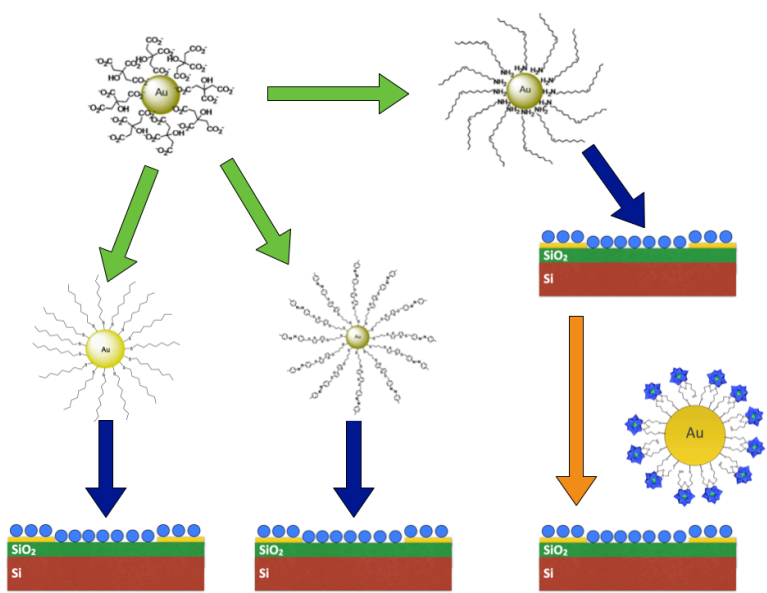

***Figure S1.*** *Scheme of the NMN synthesis routes. The green arrows stand for ligand exchange in solution, the blue arrows indicate the transfer printing method and the orange one represents the on-surface ligand exchange (see text for details). In the case of POMs, the counterions (3 TBA$^+$ per POM) are omitted for clarity*.

The next step is to form a compact 2D network of the NPs at the surface of the Si/SiO$_2$ substrate with patterned electrodes (Fig. S1, blue arrow). We used the Santhanam[6] method to form a Langmuir film at the surface of a non-miscible and non-volatile solvent. Water and ethylene glycol meet these criteria but we have obtained better quality films with ethylene glycol. In a crystallizer, we put a pierced Teflon Petri dish (hole diameter: 2 cm) upside down. We add ethylene glycol (EG) until we form a meniscus on the hole and then we spread some drops of the solution of NPs. We protect the assembly by covering with a crystallizer and wait around 10 minutes that the solvent evaporates and the film is self-organized on the EG surface. Then, we used a polydimethylsiloxane (PDMS) stamp to collect the NP films and transfer it on the surface of the SiO$_2$/electrode substrate, following the Langmuir–Schaefer technique.[7] We delicately put the PDMS stamp on the surface of the meniscus, dry the stamp under nitrogen flow and we recover the SiO$_2$/electrode substrate with this modified stamp. We take it out after a few seconds to be sure that the network of oleylamine-NPs is well transferred and we rinse quickly with ethanol the functionalized substrate. The film peels off easily from the PDMS tab. Then, we check the homogeneity and organization of the film by scanning electron microscopy (SEM), Fig.S2. We clearly observed the deposition of a monolayer of oleylamine-NPs with mainly a roughly hexagonal arrangement of the NPs and an almost homogeneous size of NPs. The zoom images were treated with ImageJ[8] to give us statistical data of the NP diameter (we used the Feret's diameter) and the inter-nanoparticle distance was calculated with the nearest neighbor distance (NND) ImageJ plugin.

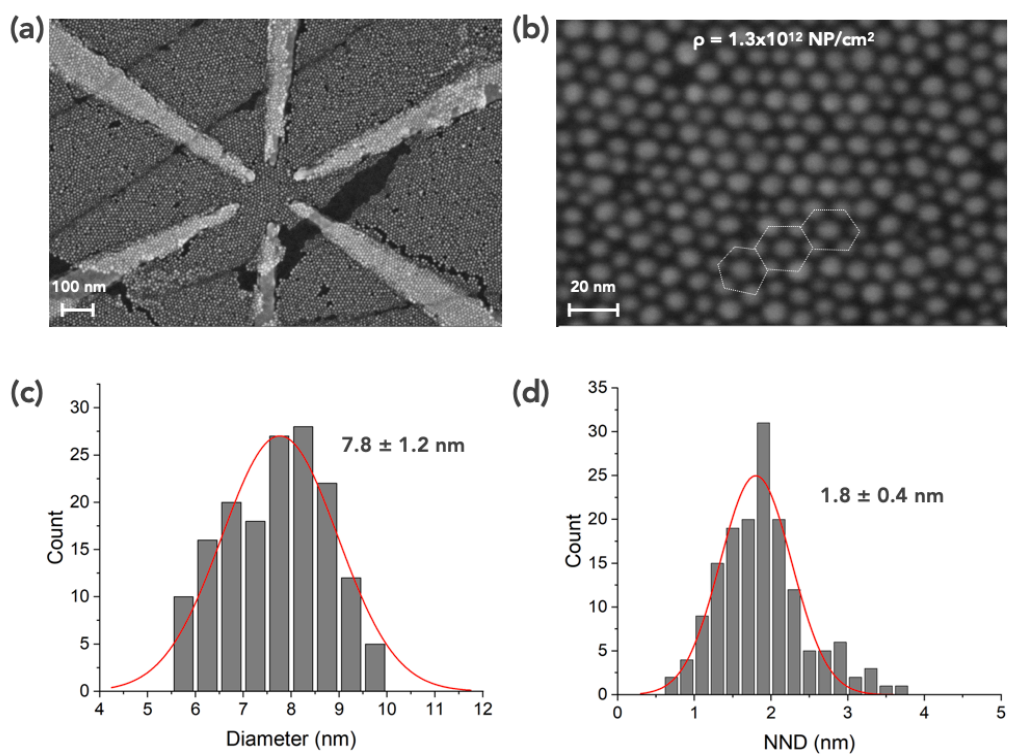

*Figure S2. (a-b) Scanning electron microscope images of the oleylamine-NMN at different magnifications (82.06k and 648.56k, respectively). The panel (a) shows the 6 electrodes and the monolayer of NPs, the central ring between the electrode has a diameter of ca. 100 nm. Panel (b) is a zoom near the electrodes. The roughly hexagonal packing of the functionalized NPs is illustrated by the dotted white lines in the panel (b). (c-d) Histograms of the NP diameter and nearest neighbor distance (NND), respectively. The red lines are Gaussian fits, the mean values and standard deviations are given in the figures.*

Figure S2 shows the distribution of the diameter of the NPs, the mean diameter is around 7.8 nm. The nearest neighbor distance (NND) is around 1.8 nm. The length of the oleylamine is ca. 2.0 nm[9] indicating that the ligands are strongly interdigitated and folded (likely at the double bond) in the gap between two neighboring NPs.

***Ligand exchange.*** The last step is the ligand exchange to replace the capping ligands (citrate or oleylamine) with the thiolated molecules (octanethiol, azobenzene, polyoxometalate), the thiol-ligand exchange was already demonstrated elsewhere.[10, 11] Transfer of citrate-NPs in organic medium was necessary for the thiolation reaction with octanethiol (Fig. S1). To this end, the 100 mL citrate-NPs solution was centrifuged at 13000 rpm for 30 min to eliminate the maximum of water supernatant. Then NPs were precipitated by the addition of an excess of ethanol and centrifugation at 10000 rpm for 5 min. After removal of the solvent, the black precipitate physisorbed on the centrifugation tube (attention, do not dry the precipitate!) was redispersed in 10 mL of absolute ethanol by sonication, providing a dark blue suspension immediately treated with 100 µL of octanethiol. The solution quickly turns red-purple but the thiolation is continued 24h at RT protected from air and light. The resulting black precipitate was washed 3 times with ethanol at low speed centrifugation (2000 rpm max), then redispersed by sonication in $CHCl_3$ for the preparation of MNMs.
The same method was used for the synthesis of azobenzene-NPs as already described in a previous work.[4]

We also tried to apply this method to prepare a suspension of POM-NPs in organic medium but it was not possible to obtain NMN films by the Langmuir technique. To get around the problem, for the preparation of POM-NMNs we opted for a ligand exchange method on a preformed NPs network (Fig. S1, orange arrow). To this end, we immersed the oleylamine-NMN substrate ($SiO_2$ with electrodes) in a $10^{-3}$ M solution of POM in acetonitrile during 5-10 minutes. Then the substrate was rinsed quickly with acetonitrile and it was dried under nitrogen flow.

Figure S3 shows the SEM characterization of the octanethiol-NMNs. The mean NP diameter is still 7.8 nm, but a tail at larger sizes indicates that some NPs are aggregated. The mean NND distance is 1.5 nm which indicates that the C8 alkyl chains (length of ca. 1.3 nm in their all-trans conformation) are strongly interdigitated and/or folded with the presence of gauche defects. We also observe a tail of the NND at larger sizes, which is due to presence of numerous voids in the layer as visible in the SEM images.

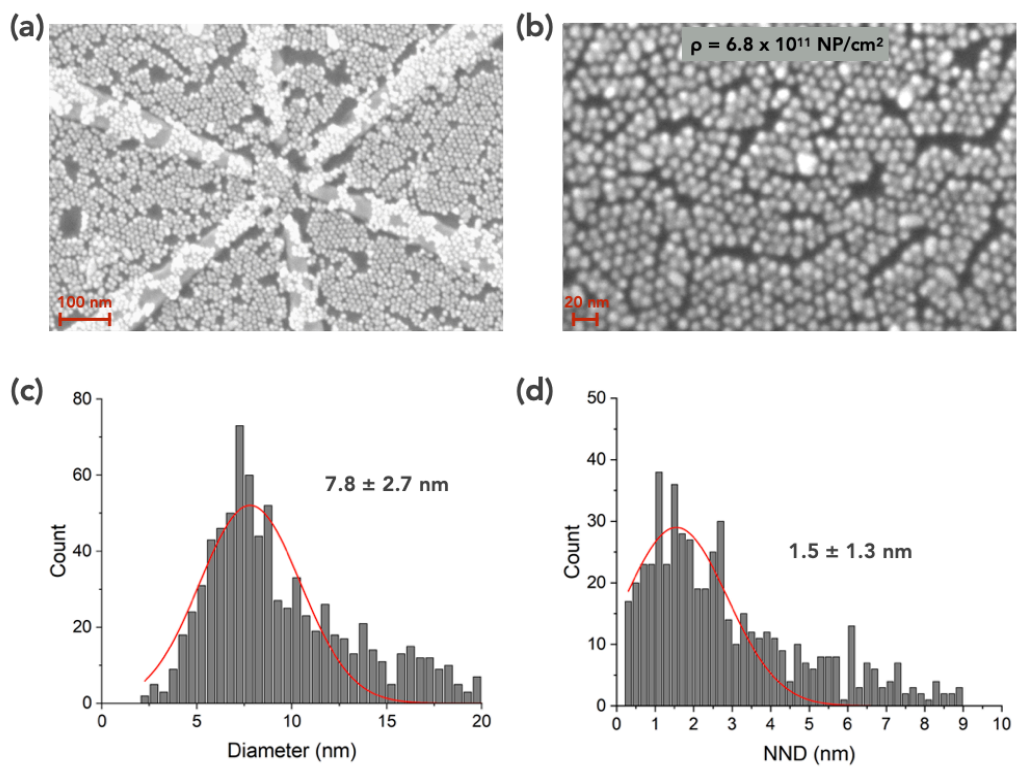

*Figure S3. (a-b) Scanning electron microscope images of the octanethiol-NMN at different magnifications (132.41k and 279.91k, respectively). The panel (a) shows the 6 electrodes and the monolayer of octanethiol-NPs, the central ring between*

*the electrode has a diameter of ca. 100 nm. Panel (b) is a zoom near the electrodes. **(c-d)** Histograms of the NP diameter and nearest neighbor distance (NND), respectively. The red lines are Gaussian fits, the mean values and standard deviations are given in the figures.*

We prepared two batches of the POM-NMNs, which have a slightly different organization of the nanoparticles (NPs) in the NMN (Fig. S4). For batch 1 the NPs are slightly denser than for the batch 2 (Fig. S4a and S4b, respectively). Figure 1 (main text) and Fig. S4a show the SEM images of the POM-NMNs after the on-surface ligand exchange (batch 1). We still have a 2D monolayer of NPs and the organization of the NPs looks stable after the exchange and we still observed a hexagonal packing. From the image analysis (Fig. S4c), the mean NP diameter is in the range 7-8 nm. The NND is slightly larger and more dispersed for the batch 2 (Fig. S4e and S4f). Compared to the size of the POM molecule (≈ 1.8 nm, see Table 1 main text, or even a bit less since the short alkylthiol legs are flexible), we assume that no more than one layer of POMs is surrounding the NPs and intercalated in the gap between two adjacent NPs.

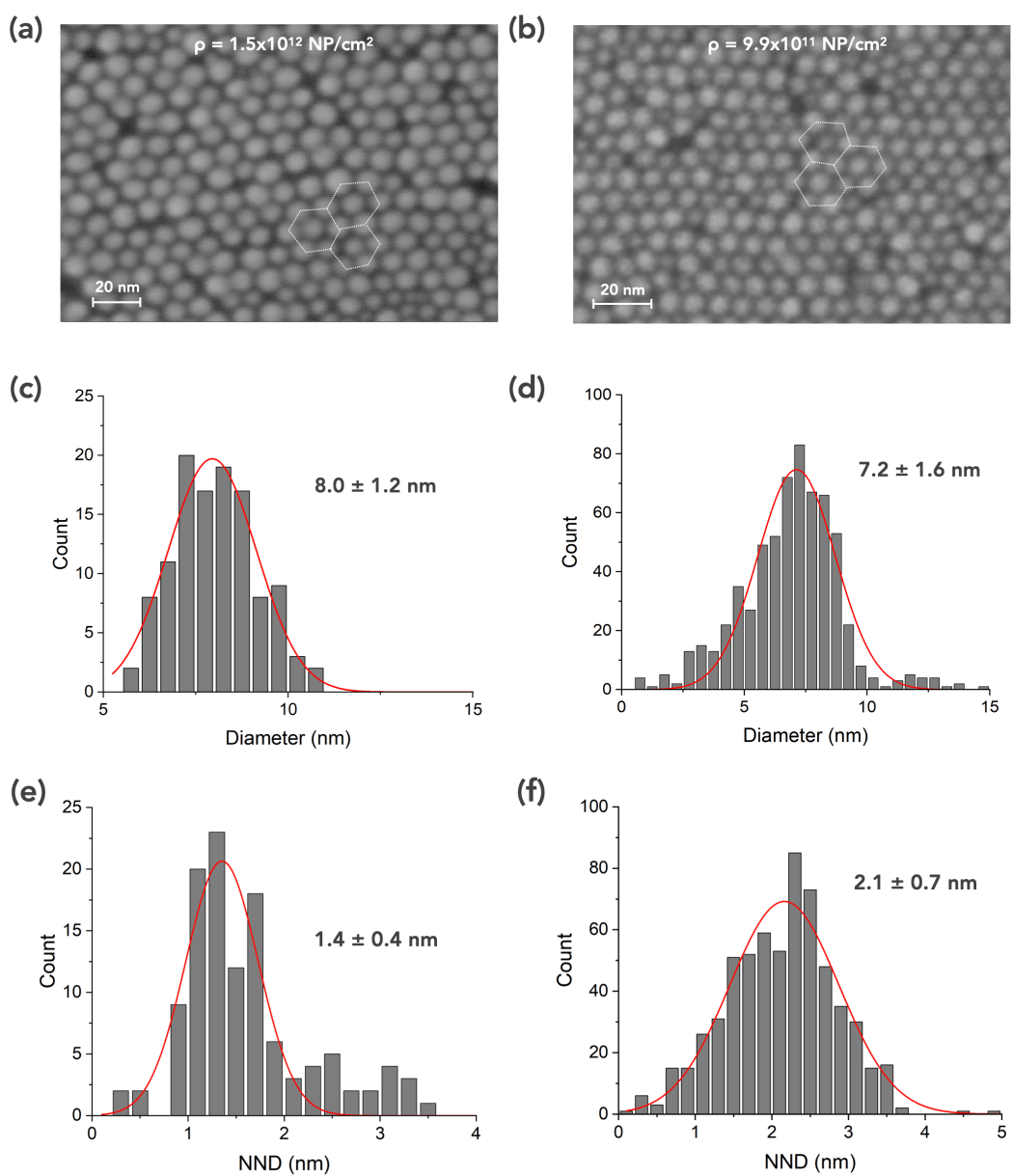

*Figure S4. (a-b)* Scanning electron microscope images of the POM-NMNs for batches 1 and 2 (magnification 652.75k and 162.96k, respectively). The hexagonal packing of the functionalized NP is illustrated by the dotted white lines in the panel. *(c-d)* Histograms of the NP diameter for the NMNs of batch 1 and batch 2, respectively. *(e-f)* Histograms of the nearest neighbor distance (NND) for *batch 1 and batch 2, respectively. The red lines are Gaussian fits, the mean values and standard deviations are given in the figures.*

For the azobenzene-NMN, the mean NP diameter was 9 nm with a mean NND of 4.5 nm as fully characterized in our previous works.[4, 12]

## Estimation of the voltage inside a single AuNP-molecule-AuNP

The voltage across an individual NP–molecules–NP building block junction in the NMNs is roughly approximated by the applied voltage divided by the number of such junctions in series between the PEs (≈5 to 15 as estimated from the SEM images for the NMN with a central diameter of ≈ 100 nm and depending on whether the PEs are diametrically located of side-by-side). For a crude estimate, we can consider that on average *ca.* one tens of the external applied voltage is sustained by single NP-molecule-NP building block.

In the case of the azobenzene-NMNs (see Fig. 3a in Ref. 12), in the voltage range 8-12 V (voltages at which the LFN behavior of the *trans* azobenzene-NMNs tends to be similar to the one of the *cis* azobenzene-NMNs) and considering an average inter-nanoparticle distance of ≈ 4.5 nm (Table 1, Ref. 4), the electric field in an individual NP–azobenzene–NP junction is ≈ 1.2-5.3x10$^{-2}$ V/Å.

**Additional data.**

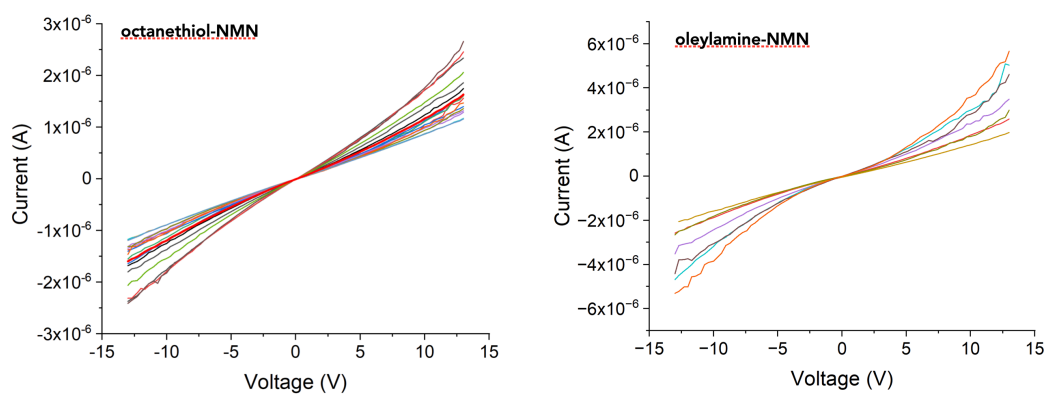

*Figure S5. Current-voltage (I-V) curves recorded for several pairs of electrodes (PEs) of the octanethiol-NMN and oleylamine NMN.*

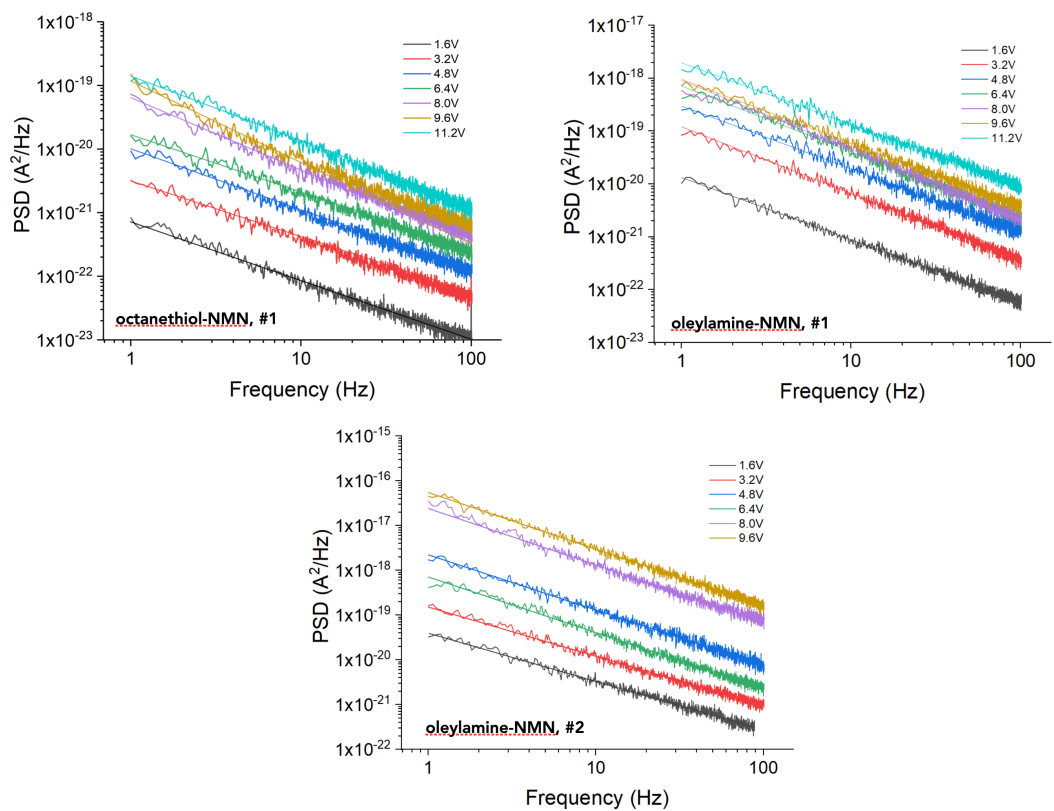

*Figure S6. Current power spectral density (PSD), $S_I(f)$, versus frequency for the octanethiol-NMN, PE #1, and the two PEs of the oleylamine-NMN, measured at several applied voltages.*

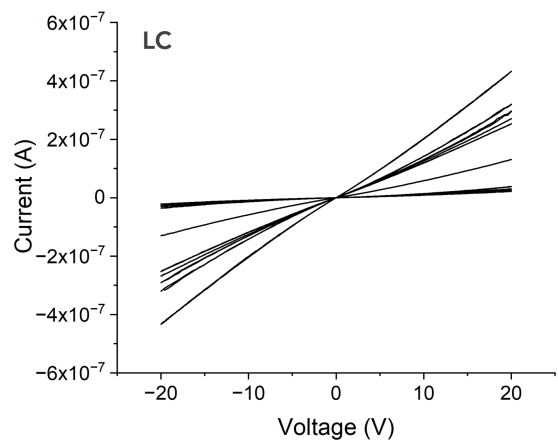

*Figure S7. Current-voltage traces of the LC POM-NMN (zoom on data from Fig. 3B, main text)*

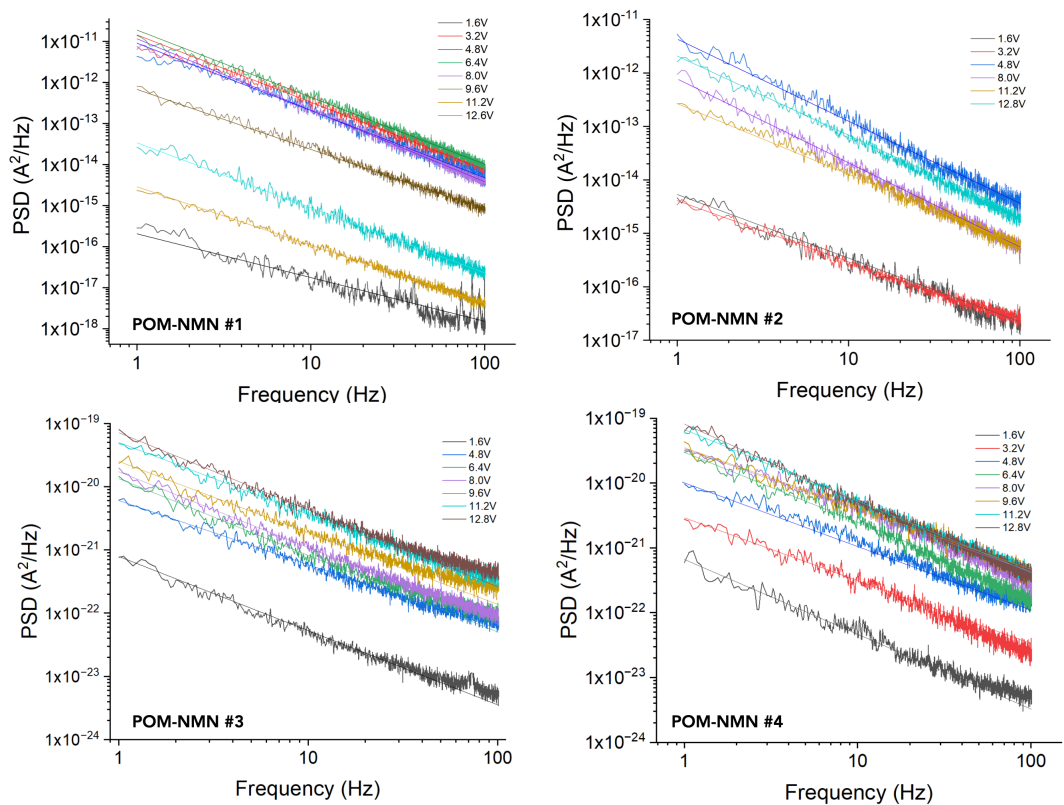

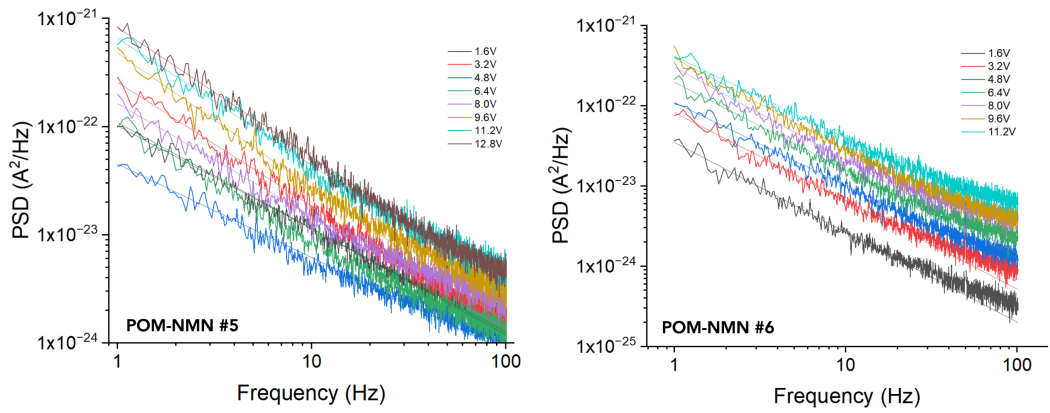

*Figure S8.* Current power spectral density (PSD), $S_I(f)$, versus frequency for the POM-NMNs (batch 1: PEs #1 and #2; batch 2: PEs #3 to #6) measured at several applied voltages.

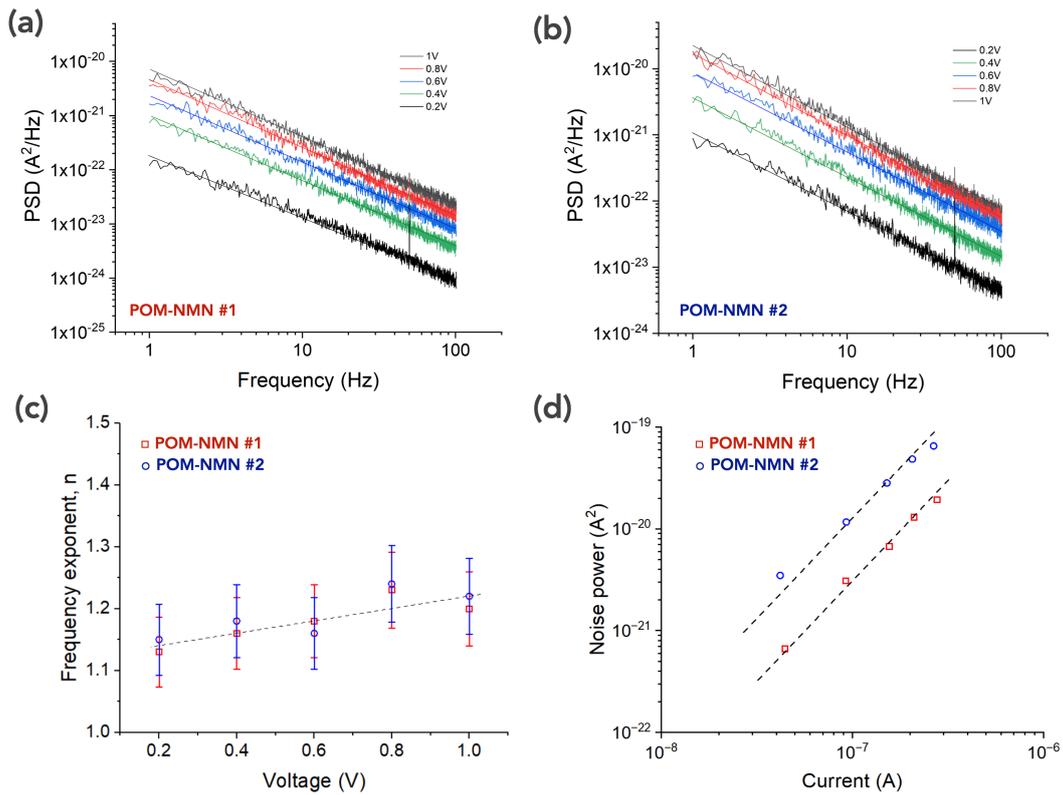

*Figure S9. (a-b)* Current power spectral density (PSD) $S_I(f)$, versus frequency for the two PEs of the POM-NMNs measured at several applied voltages from 0.2 to 1V. **(c)** *Frequency exponent, n, versus the applied voltage. The dashed line is a guide for eyes.* **(d)** *Noise power versus the DC current. The dashed lines are a guide for the eyes.*

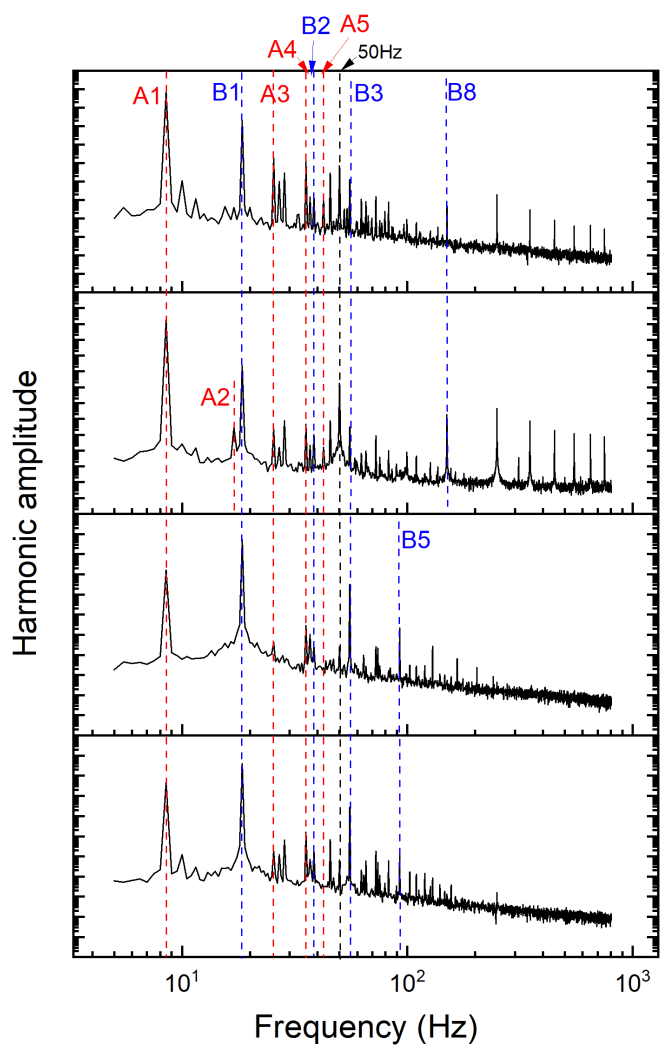

**Figure S10**. *Two sinusoidal signals, signal A at 8.5 Hz and signal B at 18.5 Hz (peak-to-peak amplitude $V_{PP}$ = 2 V for both) are applied at two electrodes of the POM-NMN. At the other 4 outputs, the currents are measured by a transimpedance amplifier and fed to the dynamic signal analyzer for FFT analysis. The HHG peaks are labeled as Ai (i = 1 for the fundamental, i = n for the $n^{th}$ harmonic, n is an integer) and Bi for harmonics corresponding to the A and B*

*input signals, respectively. Only the main HHG are shown for illustration. Peaks in between these integer harmonics correspond to interharmonic distortion and intermodulation distortion (see Ref. 12 for details on the method and analysis procedure). The large number of generated harmonics by the NMNs is the fingerprint of its strongly nonlinear response. The HHG spectra are also different for the 4 outputs, indicating the variability of the building blocks and interactions in the NMN.*

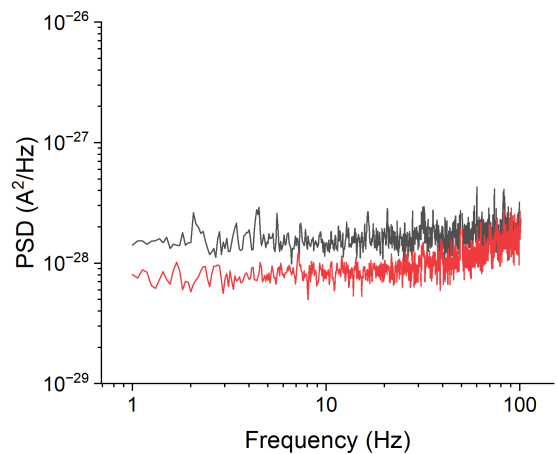

**Figure S11.** *Noise floor of the setup measured for the two channels (the two trans-impedance amplifiers and the two-channel digital signal analyzer) with the prober tips raised (no sample).*

### Reservoir computing.

The concept of reservoir computing (RC) has emerged at the beginning of the 2000s with two seminal publications of Jaeger et al.[13] and Maass et al.[14] RC is a peculiar type of the recurrent neural network and it is appropriate for temporal/ sequential information processing.[13]

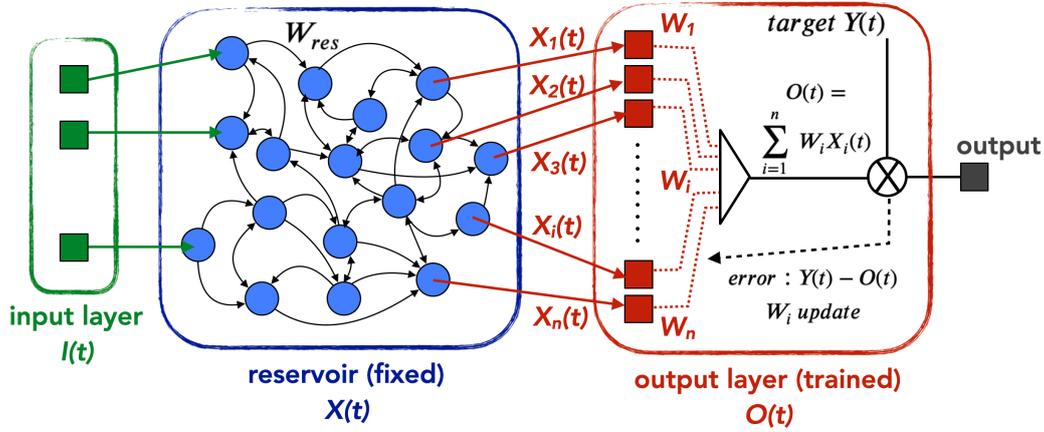

***Figure S12***. *Principe scheme of a RC that is made of an input layer, a reservoir and a trained output layer. The reservoir is a randomly interconnected (black lines) network of nodes (blue circles). The transfer functions of the links are characterized by weights $W_{res}$ that are held fixed. Several $X_i(t)$ reservoir outputs are read and weighted ($W_i$) as a linear combination to generate and output $O(t)$, which is compared to the target $Y(t)$. The error $Y(t)-O(t)$ is minimized by updating the weights $W_i$ using a learning algorithm.*

In the RC approach (Fig. S12), the time-varying input signals I(t) feed a reservoir that is characterized by complex dynamics and highly non-linear properties. The reservoir is composed of nodes (blue circles) and links (black arrows). The signals propagate between nodes that are interconnected by links with random weights $W_{res}$ that are characterized by a large variability of values. The reservoir dynamics and non-linearity generate states $X_i(t)$ that are a function of the inputs I(t), the most recent state $X_i(t-1)$ and the weights $W_{res}$ following:[13]

$$X_i(t) = f\left(I(t) + W_{res}X_i(t-1)\right) \tag{S1}$$

In the reservoir, the input signals are projected into a higher-spatio-dimensional representation space. The signals of some output nodes are read by an output layer (basically a simple perceptron)[15], where the signals are linearly weighted ($W_i$) to generate the time series output O(t) according to:

$$O(t) = \sum_{i=1}^{n} W_i X_i(t)$$

(S2)

The output layer is trained to perform a given information processing task by comparing O(t) with the target signal Y(t) and updating the weights $W_i$ with an appropriate learning algorithm. Contrary to multi-layer feed-forward neural networks and/or convolution neural networks where all the hidden layer weights need to be trained and adjusted, RC is a simplified computation system at the hardware level because only the output layer weights $W_i$ must be trained, while the reservoir weights ($W_{res}$) remain fixed. Thus, the implementation of hardware RC was tested using a variety of physical devices and technologies (see a review in Ref. 16) including nanoscale materials and devices (see a specific review in Ref. 17). One of the mandatory conditions for an efficient RC is a large variability of the $W_{res}$ values.[13, 14, 16, 18] Similarly the topology in the reservoir is fixed and random, and the transfer function of the links and nodes in the reservoir has to be strongly non-linear with a complex dynamic behavior.[18-20] Note that the output layer can be implemented physically or most of the time by a software algorithm. In the present case, only the reservoir layer has been implemented, the nodes are the gold NPs and the links are the molecules connecting neighboring NPs.